\documentstyle[12pt]{article}
\newcommand{\bee}{\begin{equation}}
\newcommand{\ee}{\end{equation}}
\newcommand{\ba}{\begin{array}}
\newcommand{\ea}{\end{array}}
\newcommand{\bea}{\begin{eqnarray}}
\newcommand{\eea}{\end{eqnarray}}
\begin{document}
\thispagestyle{empty}
\begin{flushright}
February 1995
\end{flushright}
\bigskip\bigskip
{\it Comment on}
\begin{center}
{\bf \Large{`Asympotic Scaling in the Two-Dimensional}}\vskip2mm
{\bf \Large{$O(3)$ $\sigma$-Model at Correlation Length $10^5$'}}
\end{center}
\vskip 1.0truecm
\centerline{\bf
Adrian Patrascioiu}
\vskip5mm
\centerline{Physics Department,
University of Arizona, Tucson, AZ 85721, U.S.A.}
\vskip5mm
\centerline{and}
\vskip5mm
\centerline{\bf Erhard Seiler}
\vskip5mm
\centerline{Max-Planck-Institut f\"ur
 Physik}
\centerline{ -- Werner-Heisenberg-Institut -- }
\centerline{F\"ohringer Ring 6, 80805 Munich, Germany}
\vskip 2cm

In their recent letter \cite{Caracc} Caracciolo et al claim to have
measured the temperature dependence of the correlation length in the $2D$
$O(3)$ $\sigma$-model up to $10^5$ and to find excellent (4\%) agreement
with the Hasenfratz-Maggiore-Niedermayer (HMN) formula \cite{HN}.
Their results come from applying finite size scaling (FSS) to Monte Carlo
(MC) data taken on lattices of linear size $L\leq 512$, 200 times smaller
than the alleged correlation lengths. Although this fact alone casts
doubt upon such claims, we would like to repeat here why procedures of
the type employed by Caracciolo et al cannot be used to study the
question of asymptotic scaling in $2D$ $O(N)$ models (see also
\cite{FF,Kimcom}).

FSS is clearly an asymptotic statement about the limit $L\to\infty$ at
$x=\xi(L)/L$ fixed. This means that if asymptotic scaling would hold,
$L$ would have to be increased like $O(e^{2\pi\beta})$; if, as we believe,
there is a critical point at a finite value of $\beta$, $L$ has to
increase with $\beta$ even faster.
A crucial question is how large an $L$ one should choose
if one wishes the corrections to FSS to be smaller than some given
percentage. While there is no easy answer to this question, one
criterion is provided by perturbation theory (PT).
PT is sure to provide the correct asymptotic expansion at
$L$ fixed, $\beta\to\infty$ and suggests that if
$L<<O(e^{\pi\beta})$, any MC measured quantity will be accurately
reproduced by PT.
Moreover, at fixed $L$, the accuracy with which PT reproduces MC data
will increase with increased $x$. We have verified this explicitely
in ref.\cite{Kimcom} and so have Caracciolo et al as seen in their Fig.2,
where for $x>0.7$, the PT prediction is indistinguishable from the MC data.
So their statement that they do not assume asymptotic scaling is
misleading: implicitly they do, by working in the perturbative regime
for the crucial large $x$ values.

There is another, related, trouble with PT at fixed $L$: as
we have shown explicitely \cite{si}, the two limits $L\to\infty$ and
$\beta\to\infty$ cannot be interchanged. The problem is this:
if through their FSS procedure Caracciolo et al did determine the
true $\xi_\infty^{(2)}(\beta)$, the result
should be independent of the boundary conditions (b.c.) used.
In \cite{si} we showed explicitely that in the non-Abelian
models $O(N)$ $N\geq 3$, the termwise limits of the PT
coefficients and even of the so called universal coefficients of the
$\beta$-function depend upon the b.c..

The only safe way to avoid the pollution of the FSS predictions by the b.c.
is to work on lattices with $L>>O(e^{\pi\beta})$. In their work not
only did
Caracciolo et al not obey this criterion, but in fact for $x>0.7$ they reduced
$L_{min}$ from 128 to 64. They state that they needed a larger $L_{min}$
for $x<0.7$ to eliminate certain scaling violations.
In fact these scaling violations are systematic:
As can be seen in their Fig.1, for $x<0.6$
the data points taken at the same $x$ but larger $L$ (i.e.~larger $\beta$)
generally produce larger values for the scaling function $F_\xi$
These non-perturbative scaling violations shift to larger $L$
values with increasing $x$
and are for $x>0.7$ no longer visible in the limited range of $L$ values
studied. But this certainly cannot taken as proof that the limit
$L\to\infty$ has been reached. On the contrary, it would be worth
some effort to study these scaling violations in
more detail.

Those FSS violations are also reflected in the extrapolated values
of the correlation length $\xi_\infty^{(2)}$ produced by Caracciolo et al
as well as those of Kim, reported their Tab.2:
one can see that generally larger lattices lead to larger
values of $\xi_\infty^{(2)}$.

In their letter Caracciolo et al state that while their work establishes
FSS for $L_{min}\leq L\leq 256$ and $1.65\leq\beta\leq 3$, they cannot
rule out the possibility that this situation could change for $L>256$
and for $\beta>3$, because `at some large correlation length
($\geq 10^3$) the model crosses over to a new universality
class'. This statement is incorrect: as said
above, all their results employing FSS data with $x>0.7$ are
perturbative, hence in principle polluted by b.c. effects and cannot
be regarded as true determinations of $\xi_\infty^{(2)}(\beta)$.
This means that contrary to their claim implicit in Tab.2, one does
not know $\xi_\infty^{(2)}(\beta)$ for $\beta>1.9$; in particular one
does not know if it varies in agreement with asymptotic scaling nor that
it does not diverge for $\beta<3$.

In support of their findings, Caracciolo et al invoke the improved agreement
between their $\xi_\infty^{(2)}(3.0)$ and the HMN formula. We find this claim
also misleading: if one accepts their premise that for $x>0.7$ one
can use $L_{min}=64$, then one can use PT (their eq.(7)) to compute the FSS
curve to arbitrarily large values of $x$. This way one can extend their
procedure to larger $\beta$ and see if the agreement with the HMN prediction
improves. Since at those large $\beta$ values PT at fixed $L$ becomes
very good, one can use instead of the non-existing MC data at very large
$\beta$
the PT values, use the PT from of the FSS curve to estimate $\xi_\infty^{(2)}$
and check if the agreement with the HMN prediction improves. This is
not the case, so the good agrement found at $\beta=3$ has to be considered
as accidental.

\end{document}